\documentclass[nolinenumbers,trackchanges,default]{aastex701}
\usepackage{graphicx}
\usepackage{lipsum}
\usepackage{amsmath}
\begin{document}
\nolinenumbers

\title{Rare Near-Opposition Alignment of 3I/ATLAS on 22 January, 2026}

\author{Mauro Barbieri}
 \email{mauro.barbieri@pm.me}
	\affiliation{INAF - Padova Observatory, Vicolo dell'Osservatorio 5, 35122, Padova, Italy}
   \author{Abraham Loeb}
 \email{aloeb@cfa.harvard.edu}
	\affiliation{Astronomy Department, Harvard University, 60 Garden St., Cambridge, MA 02138, USA}

%\date{\today} 

%\maketitle

\begin{abstract}
We point out that on 22 January 2026, the interstellar object 3I/ATLAS will align to within an exceptionally small angle, $\alpha = 0.69^\circ$, with the Earth-Sun axis. This rare alignment provides unique circumstances for measuring the opposition surge and polarimetric properties of interstellar cometary dust. We characterize the alignment geometry, outline key scientific opportunities, and define the observational requirements for data collection. Observations before and after the alignment time offer an unprecedented opportunity which may not repeat for decades, for characterizing the albedo, structure, and composition of interstellar matter.
\end{abstract}

\section{Introduction}

On July 1 2025, the Asteroid Terrestrial-impact Last Alert System (ATLAS) discovered the interstellar object 3I/Atlas \citep{Bolin,Selig}. Follow-up, as well as pre-discovery, observations validated its hyperbolic orbit with eccentricity $e\approx 6.139$ and perihelion distance of $q\approx 1.356\,\rm{AU}$ \citep{3AtlasJPL}, confirming its interstellar origin. Its interstellar velocity relative to the Sun of $v_\infty\approx 57.7\,\rm{km\,s}^{-1}$ is large in comparison to the two other documented interstellar objects, 1I/`Oumuamua ($v_\infty\approx26.4\,\rm{km\,s}^{-1}$, \citep{1IJPL}) and 2I/Borisov ($v_\infty\approx32.3\,\rm{km\,s}^{-1}$, \cite{2IJPL}). 

Interstellar objects provide unique opportunities for studying materials from other stellar systems \citep{SJ}. Unfortunately for that purpose, 1I/'Oumuamua did not display traces of gas or dust around it \citep{Trilling} and 2I/Borisov was only observed at phase angles relative to the Sun-Earth axis of $\alpha > 16^\circ$ and never near opposition. Here, we point out that 3I/ATLAS will reach an unprecedented near-opposition alignment on 22 January, 2026 at 13:00 UTC.

\begin{figure}
    \centering
    \includegraphics[width=1\linewidth]{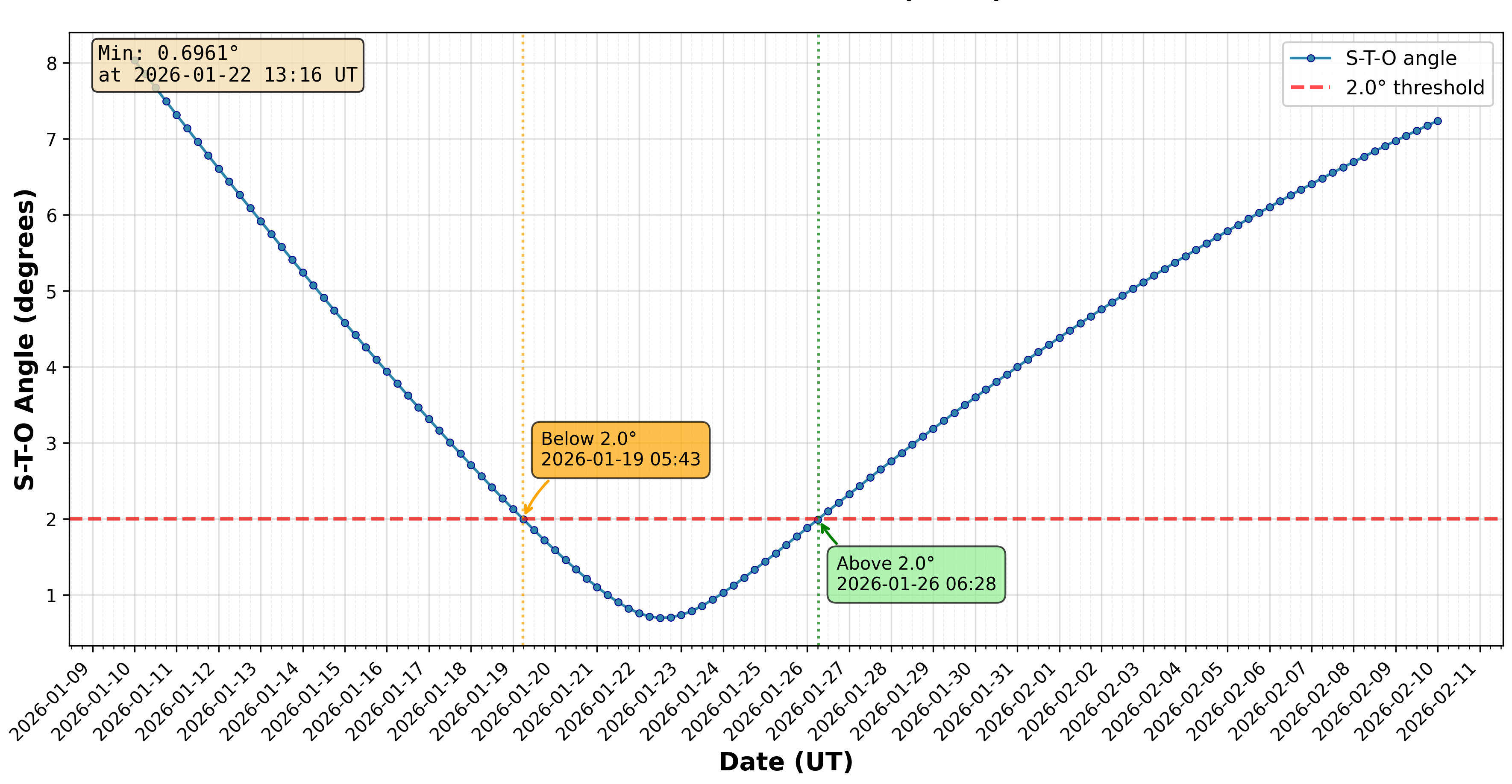}
    \caption{The angle between the Sun-Earth axis and the Sun-3I/Atlas axis in degrees, as a function of date.}
    \label{f1}
\end{figure}

At that rare time, Earth will pass nearly between the Sun and 3I/ATLAS. The phase angle $\alpha$ between the Sun-3I/ATLAS axis and the Sun-Earth axis, will reach a value of $0.69^\circ$. Unlike typical cometary opposition geometries which often last for hours, 3I/ATLAS will maintain $\alpha < 2^\circ$ for approximately one week, between 19-26 of January, 2026. Figure \ref{f1} shows the evolution of the Sun - 3I/ATLAS - Earth angle during that period, as derived from the JPl Horizons ephemeris \citep{3AtlasJPL}.

Based on JPL Horizons solution JPL\#46 (2026-Jan-08), 3I/ATLAS will be on January 22, 2026
at a heliocentric distance of $r = 3.33$ AU, a geocentric distance of $\Delta \approx 2.35$ AU, and have a V-band magnitude of $V \approx 16.7$ mag. The phase angle of 3I/ATLAS will remain small in subsequent years as it recedes from the Sun, but its magnitude will be fainter -requiring larger telescope apertures. For example, in January 2027: $\alpha \approx 1.4^\circ$, $r \approx 16$ AU, $V \approx 24$ mag, and in January 2028: $\alpha \approx 0.8^\circ$, $r \approx 28$ AU, $V \approx 25$ mag.

\section{Scientific Value of the Sun-Earth-3I/ATLAS Alignment}

At phase angles $\alpha < 10^\circ$, most Solar System bodies show a nonlinear brightness increase, called the opposition surge (see \cite{Molaro} and references therein). This surge arises from two physical effects:
\begin{itemize}
    \item 
 {\bf Shadow-hiding:} ($\alpha > 2^\circ$): When the Sun, object, and observer are nearly aligned, shadows cast by dust particles are hidden behind the particles. This eliminates dark areas, increasing the object's brightness.
\item
 {\bf Coherent backscatter:} ($\alpha < 2^\circ$): At very small angles, light traveling on reciprocal paths through a dusty medium interferes constructively, creating a narrow brightness spike.
\end{itemize}

The surge amplitude $\Delta m$ (magnitude change from small to large phase-angle values) is strongly influenced by the single-scattering albedo of dust grains $\omega_0$, as well as by the grain structure and packing. The angular scale of the surge-width constrains grain packing, as compact particles show narrow surges with half-width of order a few degrees, while fluffy fractal aggregates show broad surges with half-width of order tens of degrees.

\subsection{Current State of Knowledge}

As of now, only one comet has a well measured opposition surge: 67P/C-G. The surge was observed from the Rosetta spacecraft at $\alpha = 1.3^\circ \to 5^\circ$, yielding $\Delta m = 0.15 \pm 0.02$ mag and a very dark albedo with $\omega_0 = 0.034 \pm 0.007$ \citep{Ros17}.

For most of the solar system comets, the  opposition surge measurements are unavailable or either incomplete because of a large value for the minimum $\alpha$. The previous interstellar comet 2I/Borisov was never observed below $\alpha = 16^\circ$, far outside the opposition surge regime.

\section{Recovery of Novel Information}
 
Cometary dust is processed through its parent proto-planetary disk, and so its microphysical structure might be different from interstellar dust \citep{Draine}. The opposition surge amplitude and width of 3I/ATLAS could address the following questions:

\begin{itemize}
\item  {\bf Composition:} Is the dust shed by 3I/ATLAS dominated by carbonaceous material (low albedo, $\omega_0 \sim 0.03$) or does it retain significant ice fragments (high albedo, $\omega_0 \sim 0.1$--$0.3$), as suggested from its extended anti-tail \citep{Keto,Keto2}?

\item {\bf Grain Structure:} Are the grains compact (thermally processed) or fluffy fractal aggregates (pristine molecular cloud material)?

\end{itemize}

\section{Observational Recommendations}

The unique near-opposition geometry of 3I/ATLAS on 22 January, 2026 provides a narrow but well-defined observational window. To maximize the scientific return of community observations, we recommend the following:

\begin{itemize}
\item \textbf{Temporal coverage:} Observations should be obtained over a time span of at least $\pm$ 4 days around 22 January 2026, when the phase angle remains below $2^\circ$. This extended coverage allows separation of phase-angle effects from intrinsic activity variability.

\item \textbf{Photometry:} High-precision relative photometry ($\lesssim$0.03 mag per data point) is required to detect and characterize the nonlinear phase dependence associated with the opposition surge. Consistent aperture sizes and background subtraction methods should be used throughout the observing campaign.

\item \textbf{Multi-band observations:} Photometry in at least three broadband filters (e.g., $BVR$, $VRI$, $gri$, $riz$) would be of great value. The wavelength dependence of the phase curve provides critical diagnostics to distinguish between the shadow hiding and coherent backscatter mechanisms.

\item \textbf{Polarimetry:} Linear polarimetric measurements near minimum phase angle would offer a powerful and independent constraint on dust grain structure and multiple scattering effects. Even sparse polarimetric sampling would significantly enhance the interpretation of photometric data. New data may also help to explain the anomalous polarization properties of 3I/ATLAS \citep{Pola}.

\item \textbf{Aperture considerations:} Given the expected brightness ($V \sim 16.5$--17 mag near opposition), telescopes with apertures $\gtrsim$1~m are well suited for precise photometry, whereas larger apertures are required for polarimetric measurements.

\end{itemize}

Coordinated observations from multiple sites are encouraged to improve temporal sampling and to mitigate weather-related data gaps. Even partial datasets will contribute meaningfully to constraining the phase-angle behavior of the rare alignment of 3I/ATLAS with the Earth-Sun axis.

\bigskip
\noindent
{\bf Aacknowledgement.} A.L. is supported in part by Harvard's Black Hole Initiative and the Galileo Project.

\bibliography{literature}{}

@ARTICLE{3AtlasJPL, 
title= {{C/2025 N1 (ATLAS) Small-Body Database Lookup}}, url={https://ssd.jpl.nasa.gov/tools/sbdb_lookup.html#/?sstr=3I},
year={2025},
journal={JPL Small-Body Database},
note={{Accessed: 2025-XX-XX}},
author={Davide Farnocchia}}

@ARTICLE{2IJPL, 
title= {{C/2019 Q4 (Borisov) Small-Body Database Lookup}}, url={https://ssd.jpl.nasa.gov/tools/sbdb_lookup.html#/?sstr=2I},
year={2025},
journal={JPL Small-Body Database},
note={{Accessed: 2025-XX-XX}},
author={Davide Farnocchia}}

@ARTICLE{1IJPL, 
title= {{'Oumuamua (A/2017 U1) Small-Body Database Lookup}},
url={https://ssd.jpl.nasa.gov/tools/sbdb_lookup.html#/?sstr=1I},
journal={JPL Small-Body Database},
year={2025},
note={{Accessed: 2025-XX-XX}},
author={Davide Farnocchia}}

@ARTICLE{SJ,
       author = {{Jewitt}, David and {Seligman}, Darryl Z.},
        title = "{The Interstellar Interlopers}",
      journal = {\araa},
         year = 2023,
        month = aug,
       volume = {61},
        pages = {197-236},
          doi = {10.1146/annurev-astro-071221-054221},
archivePrefix = {arXiv},
       eprint = {2209.08182},
 primaryClass = {astro-ph.EP},
       adsurl = {https://ui.adsabs.harvard.edu/abs/2023ARA&A..61..197J}
}

@ARTICLE{Trilling,
       author = {{Trilling}, David E. and {Mommert}, Michael and {Hora}, Joseph L. and {Farnocchia}, Davide and {Chodas}, Paul and {Giorgini}, Jon and {Smith}, Howard A. and {Carey}, Sean and {Lisse}, Carey M. and {Werner}, Michael and {McNeill}, Andrew and {Chesley}, Steven R. and {Emery}, Joshua P. and {Fazio}, Giovanni and {Fernandez}, Yanga R. and {Harris}, Alan and {Marengo}, Massimo and {Mueller}, Michael and {Roegge}, Alissa and {Smith}, Nathan and {Weaver}, H.~A. and {Meech}, Karen and {Micheli}, Marco},
        title = "{Spitzer Observations of Interstellar Object 1I/{\textquoteleft}Oumuamua}",
      journal = {\aj},
         year = 2018,
        month = dec,
       volume = {156},
       number = {6},
          eid = {261},
        pages = {261},
          doi = {10.3847/1538-3881/aae88f},
archivePrefix = {arXiv},
       eprint = {1811.08072},
 primaryClass = {astro-ph.EP},
       adsurl = {https://ui.adsabs.harvard.edu/abs/2018AJ....156..261T}
}

@ARTICLE{Selig,
       author = {{Seligman}, Darryl Z. and {Micheli}, Marco and {Farnocchia}, Davide and {Denneau}, Larry and {Noonan}, John W. and {Hsieh}, Henry H. and {Santana-Ros}, Toni and {Tonry}, John and {Auchettl}, Katie and {Conversi}, Luca and {Devog{\`e}le}, Maxime and {Faggioli}, Laura and {Feinstein}, Adina D. and {Fenucci}, Marco and {Ferrais}, Marin and {Frincke}, Tessa and {Gillon}, Michael and {Hainaut}, Olivier R. and {Hart}, Kyle and {Hoffman}, Andrew and {Holt}, Carrie E. and {Hoogendam}, Willem B. and {Huber}, Mark E. and {Jehin}, Emmanuel and {Kareta}, Theodore and {Keane}, Jacqueline V. and {Kelley}, Michael S.~P. and {Lister}, Tim and {Mandt}, Kathleen and {Manfroid}, Jean and {Mar{\v{c}}eta}, Du{\v{s}}an and {Meech}, Karen J. and {Amine Miftah}, Mohamed and {Morgan}, Marvin and {Oca{\~n}a}, Francisco and {Pe{\~n}a-Asensio}, Eloy and {Shappee}, Benjamin J. and {Siverd}, Robert J. and {Taylor}, Aster G. and {Tucker}, Michael A. and {Wainscoat}, Richard and {Weryk}, Robert and {Wray}, James J. and {Yaginuma}, Atsuhiro and {Yang}, Bin and {Ye}, Quanzhi and {Zhang}, Qicheng},
        title = "{Discovery and Preliminary Characterization of a Third Interstellar Object: 3I/ATLAS}",
      journal = {\apjl},
         year = 2025,
        month = aug,
       volume = {989},
       number = {2},
          eid = {L36},
        pages = {L36},
          doi = {10.3847/2041-8213/adf49a},
archivePrefix = {arXiv},
       eprint = {2507.02757},
 primaryClass = {astro-ph.EP},
       adsurl = {https://ui.adsabs.harvard.edu/abs/2025ApJ...989L..36S}
}

@ARTICLE{Bolin,
       author = {{Bolin}, Bryce T. and {Belyakov}, Matthew and {Fremling}, Christoffer and {Graham}, Matthew J. and {Abdelaziz}, Ahmed M. and {Elhosseiny}, Eslam and {Gray}, Candace L. and {Ingebretsen}, Carl and {Jewett}, Gracyn and {Lisse}, Carey M. and {Karpov}, Sergey and {Kilic}, Mukremin and {Ma{\v{s}}ek}, Martin and {Molham}, Mona and {Roderick}, Diana and {Takey}, Ali and {Abron}, Laura-May and {Coughlin}, Michael W. and {Hsieh}, Cheng-Han and {Noll}, Keith S. and {Wong}, Ian},
        title = "{Interstellar comet 3I/ATLAS: discovery and physical description}",
      journal = {\mnras},
         year = 2025,
        month = sep,
       volume = {542},
       number = {1},
        pages = {L139-L143},
          doi = {10.1093/mnrasl/slaf078},
archivePrefix = {arXiv},
       eprint = {2507.05252},
 primaryClass = {astro-ph.EP},
       adsurl = {https://ui.adsabs.harvard.edu/abs/2025MNRAS.542L.139B}
}

@ARTICLE{Molaro,
       author = {{Molaro}, P. and {Barbieri}, M. and {Monaco}, L. and {Zaggia}, S. and {Lovis}, C.},
        title = "{The Earth transiting the Sun as seen from Jupiter's moons: detection of an inverse Rossiter-McLaughlin effect produced by the opposition surge of the icy Europa}",
      journal = {\mnras},
         year = 2015,
        month = oct,
       volume = {453},
       number = {2},
        pages = {1684-1691},
          doi = {10.1093/mnras/stv1721},
archivePrefix = {arXiv},
       eprint = {1509.01136},
 primaryClass = {astro-ph.EP},
       adsurl = {https://ui.adsabs.harvard.edu/abs/2015MNRAS.453.1684M}
}

@ARTICLE{Keto,
       author = {{Keto}, Eric and {Loeb}, Abraham},
        title = "{The physics of cometary antitails as observed in 3I/ATLAS}",
      journal = {\mnras},
         year = 2026,
        month = jan,
       volume = {545},
       number = {1},
          eid = {staf2054},
        pages = {staf2054},
          doi = {10.1093/mnras/staf2054},
archivePrefix = {arXiv},
       eprint = {2509.07771},
 primaryClass = {astro-ph.EP},
       adsurl = {https://ui.adsabs.harvard.edu/abs/2026MNRAS.545f2054K}
}

@ARTICLE{Keto2,
       author = {{Keto}, Eric and {Loeb}, Abraham},
        title = "{A Physical Model for the Ice Coma of 3I/ATLAS}",
      journal = {arXiv e-prints},
         year = 2025,
        month = oct,
          eid = {arXiv:2510.18157},
        pages = {arXiv:2510.18157},
          doi = {10.48550/arXiv.2510.18157},
archivePrefix = {arXiv},
       eprint = {2510.18157},
 primaryClass = {astro-ph.EP},
       adsurl = {https://ui.adsabs.harvard.edu/abs/2025arXiv251018157K}
}

@BOOK{Draine,
       author = {{Draine}, Bruce T.},
        title = "{Physics of the Interstellar and Intergalactic Medium}",
         year = 2011,
       adsurl = {https://ui.adsabs.harvard.edu/abs/2011piim.book.....D}
}

@ARTICLE{Ros17,
       author = {{Masoumzadeh}, N. and {Oklay}, N. and {Kolokolova}, L. and {Sierks}, H. and {Fornasier}, S. and {Barucci}, M.~A. and {Vincent}, J.-B. and {Tubiana}, C. and {G{\"u}ttler}, C. and {Preusker}, F. and {Scholten}, F. and {Mottola}, S. and {Hasselmann}, P.~H. and {Feller}, C. and {Barbieri}, C. and {Lamy}, P.~L. and {Rodrigo}, R. and {Koschny}, D. and {Rickman}, H. and {A'Hearn}, M.~F. and {Bertaux}, J.-L. and {Bertini}, I. and {Cremonese}, G. and {Da Deppo}, V. and {Davidsson}, B.~J.~R. and {Debei}, S. and {De Cecco}, M. and {Fulle}, M. and {Gicquel}, A. and {Groussin}, O. and {Guti{\'e}rrez}, P.~J. and {Hall}, I. and {Hofmann}, M. and {Hviid}, S.~F. and {Ip}, W.-H. and {Jorda}, L. and {Keller}, H.~U. and {Knollenberg}, J. and {Kovacs}, G. and {Kramm}, J.-R. and {K{\"u}hrt}, E. and {K{\"u}ppers}, M. and {Lara}, L.~M. and {Lazzarin}, M. and {Lopez Moreno}, J.~J. and {Marzari}, F. and {Naletto}, G. and {Shi}, X. and {Thomas}, N.},
        title = "{Opposition effect on comet 67P/Churyumov-Gerasimenko using Rosetta-OSIRIS images}",
      journal = {\aap},
         year = 2017,
        month = mar,
       volume = {599},
          eid = {A11},
        pages = {A11},
          doi = {10.1051/0004-6361/201629734},
       adsurl = {https://ui.adsabs.harvard.edu/abs/2017A&A...599A..11M}
}
\bibliographystyle{aasjournalv7}

\end{document}